\documentstyle[12pt,epsfig]{article}

\def\be{\begin{equation}}
\def\ee{\end{equation}}
\def\bea{\begin{eqnarray}}
\def\eea{\end{eqnarray}}
\def\ba{\begin{array}}
\def\ea{\end{array}}
\def\a{\alpha}
\def\b{\beta}

\def\d{\delta}

\def\0{$\Gamma_0$}
\def\l{\lambda}

\title{Restricted random  walks on a  graph}

\author{F. Y. Wu \\
Department of Physics\\
 Northeastern University\\
Boston,
Massachusetts 02115, USA\\
\\
and  \\
\\
H. Kunz \\
Institut de  Physique Th\'eorique\\
Ecole Polytechnique F\'ed\'erale \\
Lausanne, Switzerland}

\begin{document}

\maketitle

\medskip

\begin{abstract}
The problem of
a restricted random walk
on  graphs which keeps track of the number of immediate reversal
steps is considered
by using 
a transfer matrix formulation.
 A closed-form expression  is obtained for the generating function of
   the number of  $n$-step  walks 
with
$r$ reversal steps for walks on any graph. 
In the case of graphs of a uniform valence,
we show
that our result has a probabilistic meaning, and
deduce explicit expressions for the generating function
in terms of the eigenvalues of the adjacency matrix.
Applications to periodic lattices and the complete graph are given.
     
  \end{abstract}
\newpage

\section{Introduction}
The problem of random walks on lattices has been of pertinent  interest in
mathematics and physics for decades \cite{review}.
Very recently, the study of random walks  
also arises in   the theory of periodic orbits 
in  quantum graphs \cite{ks}.  In that study one considers 
random walks on a graph while keeping track of aspects
of steps which are immediate reversals of the previous steps,
leading to a problem of restricted walks.
As is common in   analyses of self-avoiding walks
\cite{montroll}, the 
consideration becomes difficult because of the prohibition of
returning to previously visited sites.   
As a result, the walk problem  in quantum graphs has
remained largely unsolved.  Here we introduce and  solve a  
   restricted walk problem on graphs 
which is simpler
 and more  natural in its mathematical formulation.
It is hoped that this solution will shed light on 
  self-avoiding walks
and the yet unsolved
problem of quantum graphs.

Consider a graph $G$   of $q$ sites  (vertices) numbered from 1 to $q$.
Starting from a site $p_0$, a walker begins a  walk 
by  taking  ``steps" along the edges of $G$.
 A  step from site $p_i$ to site $p_j$ is a reversal if  it 
is taken immediately after a step from $p_j$ to $p_i$, namely, if it 
reverses the preceeding step.
 We are interested in computing  $N_{n,r}(p_0, p_n)$, the number
of $n$-step walks starting from a
site $p_0$ and ending at site $p_n$,
which may or may not coincide with
$p_0$, with $r $ $(<n)$ reversal steps.
 
Introduce the generating function
\be
Z_n(z|p_0, p_n) = \sum_{r=0}^{n-1} z^r N_{n,r}(p_0, p_n). \label{z}
\ee
Clearly, we have $Z_n(0|p_0, p_n)
=N_{n,0}(p_0,p_n)$. 
It is also clear that
$Z_n(1|p_0, p_n)$ is the total number of $n$-step walks from $p_0$ to $p_n$,
regardless whether there are reversals. 
It is also useful to introduce
\bea
W_n(z|p_0) &= &\sum_{p_n=1}^q Z_n(z|p_0, p_n) \nonumber \\
   &=&  \sum_{r=0}^{n-1} z^r M_{n,r}(p_0)  \label{walk1}
  \eea
as the generating function of {\it all} $n$-step walks  which
originate from $p_0$.   Clearly,  $M_{n,r}(p_0)$ is the number of $n$-step
walks starting from $p_0$ with $r$ reversals,
regardless of the ending site.

Let {\bf A} be the $q\times q$ adjacency matrix of $G$ with elements
\bea
A_{ij}=<i|A|j>& = &1 \hskip 1cm {\rm if\>\>} i,j {\rm\>\>are\>\>connected
\>\>by\>\>an\>\>edge} \nonumber  \\
      & = &0 \hskip 1cm {\rm otherwise}. \label{matrixa}
\eea
It is well-known \cite{bm,spitzer} that we have
\bea
Z_n(1|p_0, p_n) &= & \sum_{p_1p_2\cdots p_{n-1}}
 <p_0 | A |p_1><p_1 | A |p_2> \cdots <p_{n-1} | A |p_n>  \nonumber \\
    &=& <p_0 | A^n | p_n>. \label{z1}
\eea
Here, in (\ref{z1}) and hereafter, all summations are  taken from
$1$ to $q$ unless otherwise stated.

To compute  the generating function $Z_n(z|p_0, p_n)$
for general $z$,
  we introduce a $q\times q$ matrix
{\bf W} with elements
\be
W_{ij} = 1+(z-1)\d_{ij}, \label{matrixw}
\ee
where $\d$ is the Kronecker delta, with $z=0$ corresponding to 
the case of no reversals.
Then,  instead of (\ref{z1}), we have
the expression
\be
Z_n(z|p_0, p_n) = \sum_{p_1p_2\cdots p_{n-1}}
A_{p_0p_1}A_{p_1p_2}
\cdots A_{p_{n-1}p_n} W_{p_0p_2}W_{p_1p_3}\cdots W_{p_{n-2}p_n}.
\label{z2}
\ee
A graphical representation of the 
summand in (\ref{z2})
is depicted in Fig. 1.  Note that 
arrows indicate the sequence
of steps taken. 

\begin{figure}[htbp]
\center{\rule{5cm}{0.mm}}
\rule{5cm}{0.mm}
\vskip -1.5cm
\hskip 4.5cm
\epsfig{figure=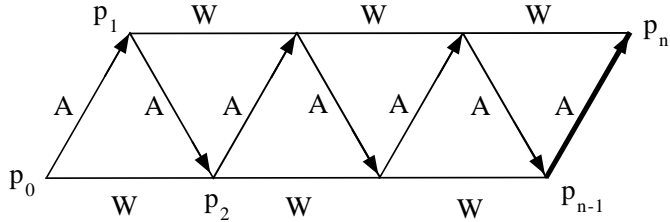,height=7.0in}
\vskip -5.3in
\caption{Graphical representation of (\ref{z2}).   Arrows denote the 
sequence of steps as well as the order
in which indices of matrix elements appear in (\ref{matrixt}).
\label{fig:fig1}}
\end{figure}

{\it Graphs of uniform valence:}
If all sites of $G$ have the same valence $v$, such as in a
complete graph or a lattice with periodic boundary conditions,
then the walker  has $v$ choices for the first step,
 and for each of the remaining $n-1$ steps,
the walker either returns  to the previous site with a weight $z$
or walks to any of the  $v-1$ remaining neighboring sites.  Then,
after summing over all ending sites, we have
\be
W_n(z|p_0) =v(z+v-1)^{n-1},  \label{walk2}
\ee
from which one obtains
\be
M_{n,r} (p_0) = v {n-1\choose r} (v-1)^{n-1}. \label{walk4}
\ee 
Furthermore, 
the generating function $Z_n$ can be associated with a
probabilistic meaning.  Consider $n$-step  walks for which the walker starts from
site $i$ and ends at $j$.  At 
the first step the walker can choose any of the $v$ neighboring sites
with an equal probability
 $v^{-1}$.
  At 
step two and thereafter, the walker makes
an immediate reversal with a probability $p_1$ and walks to any of
the remaining $v-1$ neighboring sites with
a  probability $p_2$ subject to
\be 
p_1+(v-1)p_2=1.   \label{prob}
\ee
 Then, the probability that the walker
will end at site $j$ after $n$ steps is  
\bea
P_n(i,j)& =&v^{-1} \sum_{r=0}^{n-1}p_1^rp_2^{n-1-r} N_{n,r}(i,j) \nonumber \\
  &=& v^{-1}p_2^{n-1} Z_n(p_1/p_2|i,j).\label{prob1}
\eea
Indeed, using (\ref{walk1}), (\ref{walk2})  and (\ref{prob}), one verifies  that
\be
\sum_jP_n(i,j) = 1.
\ee

\section{Transfer matrix formulation}
Returning now to walks on an arbitrary graph, it
 is  convenient to introduce
a $q^2 \times q^2$ transfer matrix {\bf T} with elements
\be
<ij|T|\ell k> = A_{ij}  W_{ik} \d_{\ell j}, \label{matrixtt}
 \ee
or more simply
\be
<ij|T|j k> = A_{ij}  W_{ik},  \label{matrixt}
 \ee
and a $q^2 \times q$ matrix ${\bf C}_0$ with elements
\be
C_0(\a \b|p_n) = A_{\a\b} \d_{\b p_n} 
\label{c0}
\ee
as depicted  by the heavy line in Fig. 1.
One can then rewrite (\ref{z2}) as
\bea
Z_n(z|p_0, p_n)&=&\sum_{p_1\cdots p_{n-1}p} <p_0p_1|{T}|p_1p_2><p_1p_2|{T}|p_2p_3>
\cdots \nonumber \\
&&\times <p_{n-2}p_{n-1}|{T}|p_{n-1}p>
C_0(p_{n-1}p|p_n).\label{z3}
\eea
Note that matrix elements in (\ref{matrixt}) and (\ref{z3}) are indexed
in the order along  the arrows   in Fig. 1.
Further introduce the recursion  relation
\be
C_{m+1}(ij|p) = \sum_{k} <ij|T|jk> C_{m}(jk|p),\hskip 1cm m=0,1,2,\cdots.
\label{c}
\ee
and the notation
\be
\a_m(i|p) = \sum_j C_m(ij|p),\label{ab}
\ee
 then for walks ending at site $p$ the recursion relation (\ref{c})
leads to
\bea
C_{m+2}(ij|p) &=& A_{ij} \biggl[ \a_{m+1}(j|p) + (z-1)C_{m+1} (ji|p)\biggr]
  \nonumber \\
          &=& A_{ij} 
\biggl[ \a_{m+1}(j|p) + (z-1)\a_{m} (i|p)+(z-1)^2 C_m(ij|p)\biggr], \nonumber \\
&&\hskip 3cm m=0,1,2,\cdots.
  \label{c1}
\eea
Here, we have iterated the first line
once  and made use of  (\ref{matrixw}) and (\ref{matrixt}) and the identity
 $A_{ij}^2=A_{ij}=A_{ji}$.
  For walks 
starting at site $i$ and ending at site $p$, we have
by combining (\ref{c0})  and (\ref{c}) the boundary condition
\bea
\a_0(i)&=& A_{ip} \nonumber \\
\a_1(i) &=& \big[{\bf A}^2\bigr]_{ip} + (z-1)v_i \d_{ip}\label{boundary}
\eea
where 
\be
v_i=\sum_j A_{ij} \label{valence}
\ee
is the valence of the site $i$.
In addition, we have the identity
 \be
Z_n(z|i,p) = \a_{n-1}(i|p).\label{z6}
 \ee
 Thus, the  crux of matter is to analyze the recursion relation 
  (\ref{c1}).

Summing (\ref{c1}) over $j$ and defining
$q$-component vectors $\vec \a_m(p)$, $m=1,2,\cdots,q, $ whose components  are $\a_{m}
(i|p), i=1,2,\cdots,q$,
we obtain
\be
\vec \a_{m+2}(p) = {\bf A} \vec \a_{m+1}(p) + {\bf B} \vec \a_m
(p), \hskip 1cm m=0,1,2,\cdots,
\label{c2}
\ee
where {\bf B} is a $q\times q$ diagonal matrix with elements
\be 
B_{ij}=(z-1)(v_i+z-1)\d_{ij}.
\ee
 Introducing the generating function
\be
\vec G(t) = \sum_{m=0} ^\infty t^m \vec \a _m(p),  \label{i}
\ee
  the summation of  (\ref{c2}) from $m=0$ to $\infty$ after multiplying by $t^m$ then
 yields
\be
\vec G(t) = \biggl[ {\bf I}-t^2{\bf B} -t{\bf A}\biggr]^{-1}
     \biggl[ \vec \a_0(p) +t \vec \a_1(p) 
    -t{\bf A} \vec \a_0(p) \biggr],
\ee
where {\bf I} is the $q\times q$ identity matrix.
Introducing (\ref{boundary}) and the vector $\vec \a_{00}(p)$ with
elements 
\be
a_{00}(i|p)  = \d_{ip}, \hskip 1cm i=1,2,\cdots,q,\label{boundary3}
\ee
one obtains 
\be
\vec G(t) = \biggl[ {\bf I}-t^2{\bf B} -t{\bf A}\biggr]^{-1}
\biggl[{\bf A} +(z-1)t {\bf V}\biggr] \vec a_{00}(p),\label{gen2}
\ee
where {\bf V} is a $q\times q$ diagonal matrix with elements
$V_{ij}= v_i \d_{ij}$.

Finally, after combining (\ref{z6}) and (\ref{i}), 
we identify the generating function
$Z_n(z|i,p)$
for $n$-step walks from $i$ to $p$  
 as the coefficient of $t^{n-1}$ in the $i$-th component of
(\ref{gen2}). This is a very general result 
which holds for any graph.

\section{Graphs of uniform valence}
In this section we apply the result of the preceeding section
to graphs of a  uniform valence, 
such as  lattices with periodic boundary conditions
and a complete graph.  
Let $v_i=v$ for all $i$, then  ${\bf B} = b {\bf I}$, 
${\bf V} = v{\bf I}$, where
\be
b=(z-1)(v+z-1),
\ee
 and the inverse matrix in  (\ref{gen2}) can be expanded to yield
\be
\vec G(t) =
 \sum_{n=0}^\infty\sum_{\ell=0}^\infty {{n+\ell} \choose \ell} 
b^\ell\ t^{n+2\ell}
  {\bf A}^n \biggl[ {\bf A}+ 
 (z-1)vt {\bf I}  \biggr] \vec a_{00}(p).\label{gen}
\ee
 Using  $\vec \a_{00}(p)$ given by (\ref{boundary3}),
we obtain after a little reduction the following explicit
expression for the $i$-th component of
$\vec G(t)$, 
\bea
&& \bigl[\vec G(t)\bigr]_i = (1-bt^2)^{-1} A_{ip} \nonumber \\
&& \hskip 0.2cm +\sum_{n=1}^\infty\sum_{\ell=0}^\infty
 b^\ell \ t^{n+2\ell}
\biggl[ {{n+\ell} \choose {\ell}} {\bf A}^{n+1}
+(z-1)v{{n+\ell-1} \choose {\ell}} {\bf A}^{n-1} \biggr]_{ip}.\label{gen3}
\eea
Thus, we find
\bea
Z_2(z|i,p) &=& \bigl[{\bf A}^2 +(z-1)v {\bf I} \bigr]_{ip} \nonumber \\
Z_3(z|i,p)&=& \bigl[ {\bf A}^3 +(z-1)(2v+z-1) {\bf A}\bigr] _{ip} \nonumber \\
Z_4(z|i,p)&=& \bigl[ {\bf A}^4  +(z-1)(3v+2z-2) {\bf A}^2 +(z-1)^2v(v+z-1)
{\bf I}\bigr] _{ip} \nonumber \\
&& {\rm etc.} \label{gen4}
\eea
It is now clear that results (\ref{gen}) and (\ref{gen4}) can 
be expressed in terms of
the eigenvalues $\lambda_i$
of the adjacency matrix {\bf A}.
 Note also that, by using (\ref{prob1}), the generating function
$Z_n(z|i,p)$  gives  the probability that a walker will arrive from
$i$ to $p$ in $n$ steps with an immediate reversal probability  $z/(1+z)$ 
at all steps.

 For walks returning to the starting point, namely, $i=p$, we use 
the identities
$A_{pp} =0, ({\bf A}^2) _{pp} = v$
to obtain
\bea
Z_2(z|p,p) &=& vz \nonumber \\
Z_3(z|p,p)&=& \bigl[ {\bf A}^3 \bigr] _{pp} \nonumber \\
Z_4(z|p,p)&=& \bigl[ {\bf A}^4 \bigr] 
_{pp} +v(z-1)(z^2-1)+v^2(z-1)(z+2).  
\eea
Furthermore, if all sites are equivalent, such as 
in a complete graph or a lattice with periodic boundary conditions,
 we have the simple relation
 \be
\bigl[{\bf A}^n\bigr]_{pp} = q^{-1}\sum_{i=1}^q \l_i^n. \label{lambda}
\ee
 For walks on an
$N_1\times N_2\times \cdots \times N_d$  $d$-dimensional
 hypercubic lattice, for
example, this becomes
\be
\bigl[{\bf A}^n\bigr]_{pp} = \bigl(N_1N_2\cdots N_d\bigr)^{-1}
\sum_{n_1=1}^{N_1} \sum_{n_2=1}^{N_2} \cdots \sum_{n_d=1}^{N_d}
\biggl[\sum_{\a=1}^d  e^{2\pi i n_\a/N_d}\biggr]^n,
\label{lambda1}
\ee
which reduces to
 the usual expression of the number of walks returning to the origin
on an infinite lattice
 \cite{spitzer} after replacing the sums by integrals in the limit
of $N_\a \to \infty, \a=1,\cdots,d$.  For the complete graph
it can be shown\footnote{For 
the complete graph
 the  problem can be 
solved very simply by regarding the walker to be in three ``states":
the starting site, just stepping off the starting 
site, and others. This leads directly to 
(\ref{new}).
 We are indebted to the referee 
for pointing out  this consideration.}  
that the solution also assumes the form 
\be
Z_n(z|i,p)= 
(q+z-2)^{n-1} (1-\d_{ip}) +(q\d_{ip} -1)\bigl[{\bf B}^{n-1} 
\bigr]_{21},
\label{new}
\ee
where
\be 
{\bf B} =\pmatrix{0 & q+z-2 & 0\cr z & 0 & q-2 \cr 1 & 0 & q+z-3},
\label{new1}
\ee
and the subscript $\{21\}$ denotes the $\{ 21 \}$-th element. 
This gives, among other results,
 the following explicit expressions for 
walks   without 
reversals at all steps,
 \bea
Z_2(0|p,p) = N_{2,0}(p,p)&=& 0 \nonumber \\
Z_3(0|p,p) = N_{3,0}(p,p) &=& (q-1)(q-2) \nonumber \\
Z_4(0|p,p) =N_{4,0}(p,p) &=& (q-1)(q-2)(q-3) \nonumber \\
Z_5(0|p,p) =N_{5,0}(p,p) &=& (q-1)(q-2)(q-3)^2 \nonumber \\
Z_6(0|p,p) =N_{6,0}(p,p) &=& (q-1)(q-2)(q^3-8q^2 +23q -23). 
   \label{walks}
\eea

\section{Summary and discussions}
We have presented a  transfer matrix formulation for
 enumerating $n$-step  walks on a graph with $r$ 
reversal steps, and obtained a
closed-form expression for the generating
function $Z_n(z|i,p)$  in the form of   (\ref{gen2}).
 We have also deduced
explicit expressions  for the generating function
in the case of   a uniform valence $v$,
with the results given in terms of the eigenvalues of the
 adjacency matrix, and showed
that in this case the generating function possesses 
a probabilistic meaning.

The transfer matrix approach can be extended  
to other restricted self-avoiding walks.  For walks
 having   a 2-step memory of not stepping into
the two immediate preceeding sites, for example, 
the generating function (\ref{z2}) can be similarly written down,
and one needs to consider  a $q^3\times q^3$ transfer matrix,
where $q$ is the number of sites in the graph.
It is hoped that such considerations can shed light to the problem
of self-avoid walks and the yet unsolved walk problems in quantum chaos.

  \section{Acknowledgement}
The work of FYW has been supported in part by National Science Foundation Grant
DMR-9614170.

\end{document}